# Biological implications of high-energy cosmic ray induced muon flux in the extragalactic shock model

Dimitra Atri[*] and Adrian L. Melott

Department of Physics and Astronomy, University of Kansas, 1251 Wescoe Hall Dr. #1082, Lawrence, Kansas 66045; atri.dimitra@gmail.com, melott@ku.edu

* Current address: Department of High Energy Physics, Tata Institute of Fundamental Research, Homi Bhabha Road, Mumbai 400 005, India

## ABSTRACT

A ~ 62 My periodicity in fossil biodiversity has been observed in independent studies of paleontology databases over ~0.5Gy. The period and phase of this biodiversity cycle coincides with the oscillation of our solar system normal to the galactic disk with an amplitude ~70 parsecs and a period ~64 My. Our Galaxy is falling toward the Virgo cluster, forming a galactic shock at the north end of our galaxy due to this motion, capable of accelerating particles and exposing our galaxy's northern side to a higher flux of cosmic rays. These high-energy particles strike the Earth's atmosphere initiating extensive air showers, ionizing the atmosphere by producing charged secondary particles. Secondary particles such as muons produced as a result of nuclear interactions are able to reach the ground and enhance the biological radiation dose. Using a Monte Carlo simulation package CORSIKA, we compute the biological dose resulting from enhanced muon exposure from cosmic rays and discuss their implications for terrestrial biodiversity variations.

1.  **Introduction**

Paleobiology databases show a ~ 62 My periodicity in terrestrial biodiversity going back ~500 My [*Rohde and Muller* 2005; *Cornette* 2007; *Lieberman and Melott* 2007; *Melott* 2008; *Melott and Bambach* 2011a,b]. Bailer-Jones [2009] has a number of criticisms; space does not allow a full discussion, but see Melott and Bambach [2011a]. He questioned inability to exclude alternative nulls, i.e. functions which appear periodic but are not. As a practical matter, one can never settle this question with a finite sample of an infinite time series. We define here (experimentally detected) periodicity as something that is significantly periodic over the available time segment. Another concern was the weakening of the signal over the last 150 Myr. This is now understood as a side effect of the declining fraction of short-lived genera, which show the signal through the full period (see also Bambach et al. 2012, submitted to Paleobiology). Also, the signal is now evident in three independent marine paleontological datasets [*Melott and Bambach* 2011a]. The correlated timing of mass extinctions appears to be consistent with random events overlaid on a periodic background stress [*Arens and West* 2008; *Feulner* 2011; *Melott and Bambach* 2011b].

The physical cause driving this biodiversity cycle is not known. One hypothesis is based on its correlation with the motion of our solar system in the galactic disk. (For an alternative, see Melott et. al 2011, submitted to Journal of Geology.) Medvedev and Melott [2007] suggested that the coincidence of period and phase of these seemingly unrelated phenomena arises through the following mechanism: The Milky Way is falling toward the Virgo cluster at ~200 km/s. A galactic shock formed at the north end of our Galaxy due to this motion is capable of accelerating particles, exposing the northern side to a higher flux of cosmic rays up to about PeV energies. The earth is protected from these cosmic rays by the galactic magnetic field when it is within the galactic disk or at the southern side. But as it moves up the galactic plane [*Gies and Helsel* 2005], the magnetic shielding is reduced and as a result it receives an enhanced flux of high-energy cosmic rays. Every ~ 62 My when we are at the north the biodiversity declines. This produces a ~10 My [*Medvedev and Melott* 2007; *Melott et al*. 2010] stress on the biosphere, accounting for diversity declines and increased severity of extinctions.

Air showers generated by the cosmic ray primaries ionize the atmosphere and can irradiate the surface with secondary radiation. Ionizing radiation can be very damaging to life. Increased atmospheric ionization leads to ozone depletion, which increases the flux of solar UVB radiation at the surface and is potentially harmful to living organisms [*Melott and Thomas* 2011]. Previously we computed atmospheric ionization [*Atri* et al. 2010a] due to extragalactic cosmic ray shock and the resulting UVB increase from ozone depletion [*Melott* et al. 2010].

Apart from ionizing the atmosphere, high-energy particles generated in the shower reach the ground and contribute to the exposure at the surface. Muons are the most penetrating ionizing radiation, dominating the dose at the ground [*O'Brien* et al. 1998; *Simonsen* et al. 2000] and the top ~1 km of ocean [*UNSCEAR* 1966]. Other components such as electrons and photons are already absorbed in the atmosphere and those reaching the surface in modest quantities are not biologically effective. Alpha particles are not penetrating and are stopped by a thin sheet of paper or by the skin.

The biological impact of cosmic rays has been studied, primarily to evaluate the damage to the human body in air and space travel [*Cucinotta* 2008]. Under normal circumstances, there is always some amount of biological damage caused by the normal flux of cosmic rays, but life has natural repair mechanisms that can repair some of the damage. If the dose is increased, this repair mechanism can be inadequate which causes dangerous mutations [*McNulty* et al. 1974] leading to various diseases, including cancer. Biological damage is quantified using experiments in which a sample is exposed to a variety of types and doses of radiation. The flux of particles at the surface is dominated by high-energy muons and other lower energy components that are less biologically effective. The impact of muons with energies greater than 10 GeV on biological samples has not been studied experimentally so far. In order to conduct experiments with such muons, one needs to tap them from particle accelerators. Higher energy muons are difficult to produce in the laboratory, so a number of theoretical studies have been conducted to calculate the conversion factor for muons on human samples. For muons, this conversion factor remains fairly constant with increasing muon energy [*Chen* 2006; *Ferrari* et al. 1997; *Pelliccioni* 2000]. Also, dE/dx for muons is a very slow function of energy [*Groom and Klein* 2000] and ionization produced by muons thus increases modestly with energy.

**2. Method**

We have calculated the secondary muon flux at the surface from high-energy cosmic ray primaries using Monte Carlo solutions to propagate air showers. (Analytical methods are quick and easy to implement but do not provide accurate estimates of the muon flux.) We used the publicly available lookup table generated by the Monte Carlo package CORSIKA [*Heck* et al. 1998] for primaries in the 10 GeV – 1 PeV range [*Atri and Melott* 2011]. CORSIKA is a widely used Monte Carlo tool, which propagates cosmic ray showers down to the ground level from primaries up to highest observable energies. The table was generated with 1.9 x $10^6$ primaries. The muon flux generated from the lookup table data is in excellent agreement with the Hebbeker and Timmermans [2002] polynomial fit which itself is derived from a compilation of muon data from a number of experiments [*Atri and Melott* 2011]. This table provides the energy distribution of muons at the surface for each primary particle averaged over the hemisphere. One convolves the cosmic ray spectrum at the top of the atmosphere with the lookup table to get the resulting muon contribution at the ground. The extragalactic shock model [*Medvedev and Melott* 2007] gives two spectra for enhanced cosmic ray exposure bracketing the uncertainties in parameters controlling cosmic ray propagation in the model, also described in detail in Melott et al. [2010]. The two spectra represent the minimum and maximum values of cosmic ray enhancement in the model, and will be referred to as Case 1 and Case 2 respectively. The normal background level of muon flux is calculated using the cosmic ray spectrum obtained from Usoskin et al. [2006].

3. **Results**

The muon flux in Case 1 shows a flux increase as well as a shift to higher energies as expected (Figure 1). The total muon energy deposition is also enhanced by 47% because of increased flux and increased average muon energy. For Case 2, there is a much higher increase in flux and the spectrum shifts to higher energies (Figure 1). The total energy deposition in this case is increased by a factor of ~16.

Of the overall globally averaged annual radiation dose from natural sources 2.4 mSv/yr, the cosmic ray component is 0.39 mSv/yr [*IAEA*]. Although the natural radiation dose has varied considerably during the past ~4 Gy period, no significant changes are known in the past ~500 My considered in this work [*Karam* 2003]. We will now compare the enhanced muon dose to the present-day natural sources. Under ordinary circumstances, muons contribute 85% to the biological dose from cosmic rays at the surface [*O'Brien* et al. 1998; *Alpen* 1998]. Therefore, the globally averaged annual dose from muons is 0.33 mSv/yr. Terrestrial sources (including radon, food, water, etc.) which are primarily from radionuclides of the uranium-thorium series, contribute 0.46 mSv/yr [*IAEA*]. The increase in the muon flux predicted by the extragalactic shock model will considerably increase the radiation dose from cosmic rays and can be potentially harmful to the biosphere.

The effective radiation dose from muons is given by:

$$D_\mu = \int \Phi_\mu \, w_{FD} \, dE$$

Where $\Phi_\mu$ is the muon flux on ground and $w_{FD}$ is the flux-to-dose conversion factor for muons. We can use the muon flux calculated above for both cases and convolve with the flux-to-dose conversion factor in order to estimate the effective radiation dose. The flux-to-dose conversion factor for human samples has been theoretically determined for muons up to 10 TeV. To calculate the biological effects over the 500 My timescale, studies on non-human samples need to be conducted. But, studies suggest that biological damage is proportional only to overall muon flux, and the fluence-to-dose factor remains fairly constant with energy [*Chen* 2006; *Ferrari* et al. 1997; *Pellicioni* 2000]. Therefore, we calculate the ratio of fluxes for muon enhancements in both cases in order to get an estimate of the biological radiation dose.

$$D_{Enhanced}/D_{Normal} = \text{Flux-Enhanced/Flux-Normal}$$

For Case 1, we saw the muon flux enhancement of 88% and for Case 2 the enhancement was by a factor of 24.5 from the normal. This would translate to an increase in the effective dose for Case 1 to 0.62 mSv/yr, which is 26% of the total radiation dose and for Case 2 to 8 mSv/yr, which is 3.36 times the present total annual radiation dose from natural sources. The total dose enhancement is by a factor of 1.26 in Case 1 and by 4.36 in Case 2. This periodic radiation dose is in addition to the normal background astrophysical radiation dose [*Karam* 2003] in the 542 My time period.

## 4. Discussion

We have computed the enhanced cosmic ray induced terrestrial muon flux predicted by the extragalactic shock hypothesis. Enhanced levels of muon flux can significantly increase mutation rates and can have significant biological implications. Aside from direct irradiation by muons, high-energy cosmic ray induced air showers leads to ozone depletion and enhanced solar UVB which is carcinogenic. We have shown that the magnitude of this effect [*Melott et al*. 2010] is small compared direct irradiation by muons. Enhanced ionization in the lower atmosphere can also lead to cloud cover changes according to some studies but the magnitude of such a change can not be determined at present due to limited progress in this field [*Atri* et al. 2010b; however see *Kirkby* et al. 2011]. We will soon compute the neutron flux and its contribution to biological damage. However, enhanced muon flux directly affects living organisms and could be the primary mechanism capable of driving biodiversity decline. Increase in the background radiation dose can result in increased mutation rate and carcinogenic diseases over a period of ~10 My, when the cosmic ray exposure from this mechanism is highest.

Increased muon irradiation will increase mutation rates, and can potentially lead to profound biological effects. It must be mentioned that these estimates provide only a rough estimate of the potential biological implications of the enhanced muon flux.
There has been no quantitative work focused on the effects of ionizing radiation at the genomic level. Such an effort will not only help better understand the relationship between radiation and biology but will also provide us with a better understanding of the evolution of life on Earth.

**5. Acknowledgements**

We thank the two anonymous referees for helpful suggestions. The extensive computer simulations were supported by NSF via TeraGrid allocations at the National Center for Supercomputing Applications, Urbana, Illinois, TG-PHY090098, TG-PHY090067T and TG-PHY090108. Research was supported in part by NASA Astrobiology grant NNX09AM85G.

**7. Figure captions**

Figure 1: Enhanced muon flux from Case 1 (dots) and Case 2 (dashes) in the extragalactic shock model compared with the normal muon flux (solid).